\documentclass[letter]{aa}  

\usepackage{graphicx}
\usepackage{txfonts}
\usepackage[colorlinks=true,citecolor=blue]{hyperref}
\usepackage{graphicx}
\usepackage{times}
\usepackage{natbib}
\usepackage{color}
\usepackage{multirow}
\usepackage{todonotes}
\usepackage{appendix}
\usepackage{array}
\usepackage{verbatim}

\begin{document} 

\title{Mapping ``out-of-the-box'' the properties of the baryons \\ in massive halos}

\author{M. Angelinelli\inst{1,2} \fnmsep\thanks{\email{matteo.angelinelli2@unibo.it}}
          \and
          S. Ettori\inst{2,3}
          \and
          K. Dolag\inst{4,5} 
          \and
          F. Vazza\inst{1,6,7}
          \and
          A. Ragagnin\inst{1,8,9}
          }

   \institute{Dipartimento di Fisica e Astronomia, Università di Bologna, Via Gobetti 92/3, 40121 Bologna, Italy
         \and
             INAF, Osservatorio di Astrofisica e Scienza dello Spazio, via Pietro Gobetti 93/3, 40121 Bologna, Italy
         \and
             INFN, Sezione di Bologna, viale Berti Pichat 6/2, 40127 Bologna, Italy
         \and
            Universitäts-Sternwarte, Fakultät für Physik, Ludwig-Maximilians-Universität München, Scheinerstr.1, 81679 München, Germany
        \and 
            Max-Planck-Institut für Astrophysik, Karl-Schwarzschild-Straße 1, 85741 Garching, Germany
         \and 
             Hamburger Sternwarte, University of Hamburg, Gojenbergsweg 112, 21029 Hamburg, Germany
         \and
            Istituto di Radio Astronomia, INAF, Via Gobetti 101, 40121 Bologna, Italy
         \and
            INAF, Osservatorio Astronomico di Trieste, via G.B. Tiepolo 11, 34143 Trieste, Italy
        \and
            IFPU, Institute for Fundamental Physics of the Universe, Via Beirut 2, 34014 Trieste, Italy
        }

   \date{Received / Accepted}


\abstract    
{We study the distributions of the baryons in massive halos ($M_{vir} > 10^{13} \ h^{-1}M_{\odot}$) in the {\it Magneticum} suite of Smoothed Particle Hydrodynamical cosmological simulations, out to the unprecedented radial extent of $10 R_{500,\mathrm c}$. We confirm that, under the action of non-gravitational physical phenomena, the baryon mass fraction is lower in the inner regions ($<R_{500,\mathrm c}$) of increasingly less massive halos, and rises moving outwards, with values that spans from 51\% (87\%) in the regions around $R_{500,\mathrm c}$ to 95\% (100\%) at $10R_{500,\mathrm c}$ of the cosmological value in the systems with the lowest (highest; $M_{vir} \sim 5 \times 10^{14} \ h^{-1}M_{\odot}$) masses. 
The galaxy groups almost match the gas (and baryon) fraction measured in the most massive halos only at very large radii ($r>6 R_{500,\mathrm c}$), where the baryon depletion factor $Y_{\rm bar} = f_{\rm bar} / (\Omega_{\rm b}/\Omega_{\rm m})$ approaches the value of unity, expected for ``closed-box'' systems. 
We find that both the radial and mass dependency of the baryon, gas, and hot depletion factors are predictable and follow a simple functional form.
The star mass fraction is higher in less massive systems, decreases systematically with increasing radii, and reaches a constant value of $Y_{\rm star} \approx 0.09$, where also the gas metallicity is constant, regardless of the host halo mass, as a result of the early ($z>2$) enrichment process. 
}   

\maketitle

\begin{figure}
\includegraphics[width=0.49\textwidth]{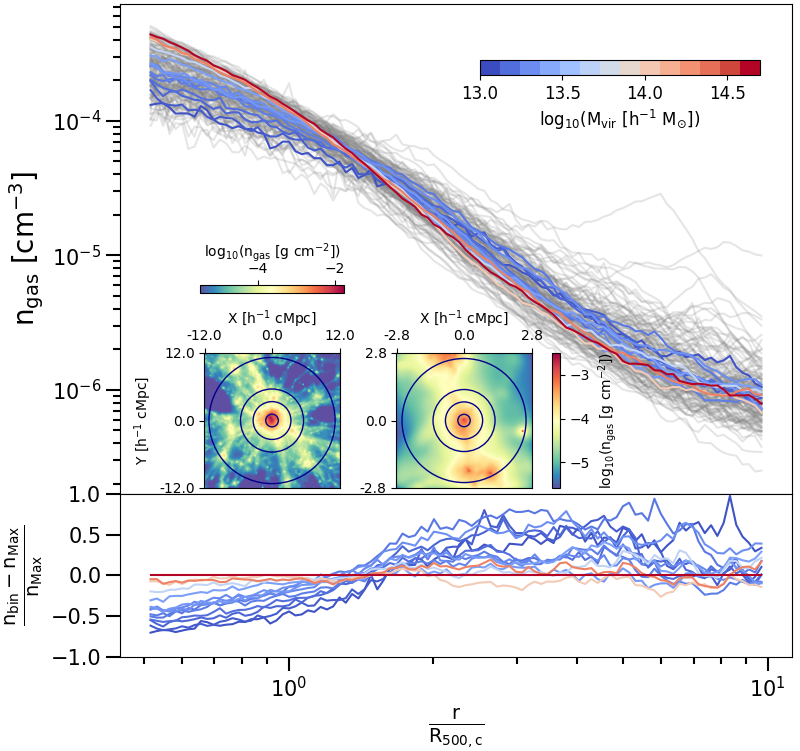}
\caption{(Top) Radial profiles of gas density from $0.5R_{500,c}$ up to $10R_{500,\mathrm c}$. Gray lines indicates profiles of single objects, while colored lines are the median values computed in different mass bins (see color legend);
(Bottom) Median radial profiles of gas density of different mass bins computed with respect to the median radial profile of the most massive bin (same color legend of the upper plot);
(Insets) Projected 2D electron gas density maps for the most (left; $M_{\mathrm vir} = 9.8\cdot10^{14} \ h^{-1}M_{\odot}$, $R_{500,\mathrm c} = 1120.4 \ h^{-1}$ckpc) 
and the less (right; $M_{vir} = 10^{13} \ h^{-1}M_{\odot}$, $R_{500,\mathrm c} = 252.5 \ h^{-1}$ckpc) massive objects in our sample. The blue circles represent 1, 3, 5 and 10 $R_{500,\mathrm c}$.
}
\label{fig:dens_maps}
\end{figure}

\section{Introduction} \label{sec:introduction}

Knowing the baryon content of groups and clusters of galaxies, the most massive gravitationally bound structures in the Universe, is key to connect their evolution with cosmology.  
A general expectation is that the relative amount of baryons they contain, with respect to their total mass, must match the ratio between the cosmological baryon density $\Omega_{\rm b}$ and the total matter density $\Omega_{\rm m}$.
However, for this to be true, sufficiently large volumes must be considered in order to treat such systems as "closed boxes" \citep{Gunn72,Bertschinger85,Voit2005}, and allow to neglect the additional, time-integrated effect of feedback from galaxy formation \citep[e.g.][]{allen11}.
Any departure from the ideal condition of a "closed box" should leave an imprint on the distribution of the different components of the baryon budget both as function of radius and halo mass \citep{Limber59}. 
Galaxy groups with typical masses of $10^{13-14} M_{\odot}$, being characterized by a shallower gravitational potential than the one found in more massive systems, can hardly be considered as "closed". They are thus unique laboratories where to study the interplay of many different physical processes affecting the evolution of baryonic matter during the hierarchical structure formation \citep{sp05}, and the energetic feedback from the many galaxies that co-evolve within them (or are continuously accreted over time). Being galaxy groups also at the peak in the halo mass function makes their cosmological and astrophysical role particularly relevant.

Many observational studies \citep{Sun09,Ettori15,Lovisari15,Eckert16,Nugent20} show how the baryon fraction in the central region ($<R_{500,\mathrm c}$) of galaxy clusters and groups increases with the mass of the system. In the external regions, the gas distribution is hardly constrained by X-rays observations because of their low signal with respect to the local background, although they are physically more interesting because of the increased complexity of processes that regulate the status of the gas, including some expected residual amount of non-thermal pressure \citep[e.g.][]{Angelinelli20}.  
If the expectations for a self-similar formation scenario is met for the most massive systems, on groups' scale the baryon content is only a half of the one expected from the self-similar scenario \citep[see e.g. review by][]{Eckert21}. 
While the low X-ray surface brightness hamper the observations to constrain the gas content of galaxy groups and clusters \citep[but for the most massive systems, constraints around $R_{200,\mathrm c}$ are available in, e.g., ][]{Ghirardini:2019}, numerical simulations are overall able to correctly reproduce many different observed physical proprieties of the gas enclosed in galaxy groups, even if the proprieties of galaxies are not well constrained
\citep[see e.g.][]{Oppenheimer21,Gastaldello21}.
Galaxy groups are the environment where the physical mechanisms that occur on galactic scale become less dominant, but still important, with respect to the gravity that rules on clusters scale. 
Numerical cosmological large scale simulations are then needed to recover the physical properties of these elusive and numerous systems \citep[][and references therein]{Ettori06,Planelles13,Haider16,McCarthy17,Pillepich18,biffi18,VallesPerez20,Galarraga-Espinosa21}, also to feed realistic prediction for next X-rays observatories generation \citep{2018A&A...618A..39R}.

In this work we perform a dedicated analysis of baryon and gas fraction on regions 
in the outskirts and beyond galaxy clusters in a sample of halos simulated in the \textit{Magneticum}\footnote{http://www.magneticum.org} suite. This work s structured as follows: in Sect.~\ref{sec:methods}, we briefly describe the \textit{Magneticum} simulations and our sample; in Sect.~\ref{sec:results}, we present the results of our analysis and the comparison with recent observational and numerical constraints. In Sect.~\ref{sec:conclusions}, we discuss our main findings, the limitations of our analysis and possible implications for future work. 

\section{Cosmological simulations: {\it Magneticum}} \label{sec:methods}

\begin{figure*}[ht]
\includegraphics[width=0.49\textwidth]{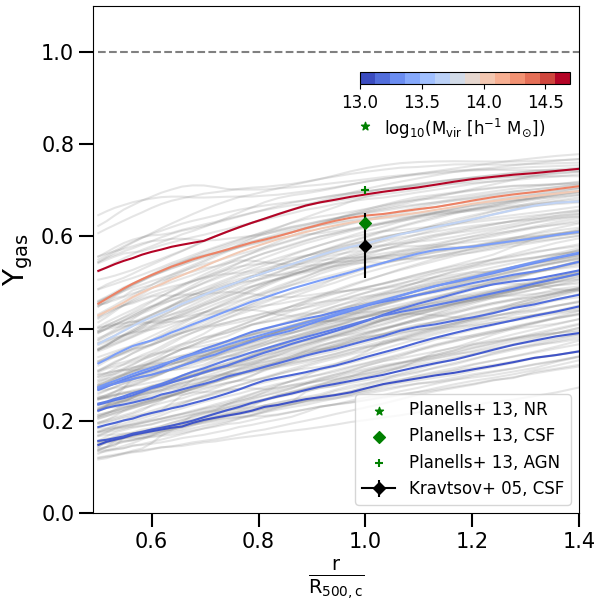}
\includegraphics[width=0.49\textwidth]{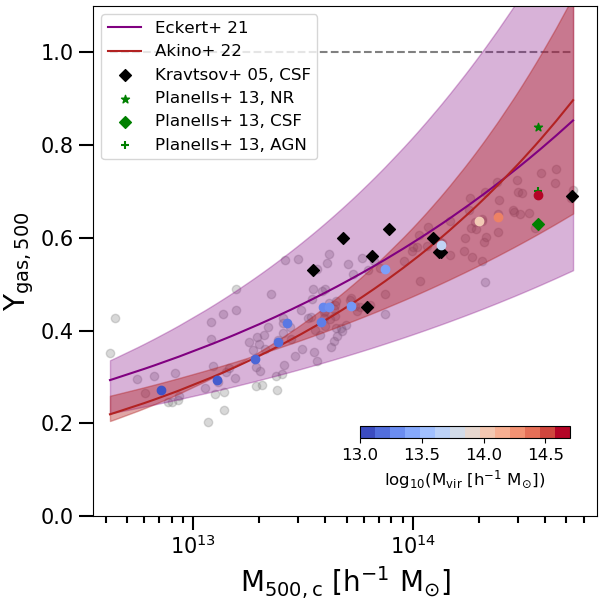}
\caption{(Left) Radial profiles of gas depletion, from $0.5R_{500,\mathrm c}$ up to $1.4R_{500,\mathrm c}$. Gray lines represent single objects in our sample, while the colored ones are the median values computed in mass bins. Comparison numerical estimates from the literature are given by different symbols  (black, Kravtsov et al. 2005; green, Planelles et al. 2013). From \citet{Kravtsov05}, we consider the mean value and its scatter for $Y_{\rm gas}$ at $R_{500,\mathrm c}$ in the simulations with gas dynamics and several physical processes (CSF). From \citet{Planelles13}, we plot the non radiative run (NR), star formation and feedback from supernova explosion one (CSF) and a run with the additional contribution from AGN (AGN).
(Right) Gas depletion parameter at $R_{500,\mathrm c}$ as function of $M_{500,\mathrm c}$. Gray dots represent single objects in our sample, while the colored ones are the median values computed in mass bins. The constraints from \citet{Kravtsov05} (black points) are indicated in correspondence of the mass of single clusters in their analysis, whereas for \citet{Planelles13} (green points) are indicated in correspondence of our most massive bin. The lines and related shadow regions are the best fit proposed by \citet{Eckert21} (purple; $f_{\rm gas,500} = 0.079^{+0.026}_{-0.025} ( M_{500} / 10^{14} M_{\odot} )^{0.22^{+0.06}_{-0.04}}$) and \citet{Akino22} (firebrick; $ln (M_{\rm gas}/10^{12} M_{\odot} ) = 1.95^{+0.08}_{-0.08} + 1.29^{+0.16}_{-0.10} ln(M_{500}/10^{14} M_{\odot})$.}
\label{fig:ygas_obs_sim}
\end{figure*}

We select a sub-sample of galaxy clusters and groups, part of the simulated \textit{Box2b/hr} of \textit{Magneticum} simulations at redshift z=0.25, the last available snapshot for this simulated box. The high resolution run of \textit{Box2b} includes a total of $2\cdot2880^3$ particles in a volume of $(640 \ h^{-1}\rm cMpc)^3$. The particles masses are $6.9\cdot10^8 \ h^{-1}\rm M_{\odot}$ and $1.4\cdot10^8 \ h^{-1}\rm M_{\odot}$, respectively for dark matter and gas component and the stellar particles have softening of $\epsilon=2 \ h^{-1}\rm ckpc$. The cosmology adopted for these simulations is the WMAP7 from \citet{2011ApJS..192...18K}, with a total matter density of $\Omega_{\rm m}=0.272$, of which 16.8$\%$ of baryons, the cosmological constant $\rm \Lambda_{0}=0.728$, the Hubble constant $\rm H_{0}=70.4 \ \rm km/s/Mpc$, the index of the primordial power spectrum $\rm n=0.963$ and the overall normalisation of the power spectrum $\sigma_{8}=0.809$. 
The more relevant physical mechanisms included in \textit{Magneticum} are:
cooling, star formation and winds with velocities of $\rm 350 \ km/s$ \citep{2002MNRAS.333..649S}; 
tracing explicitly metal species (namely, C, Ca, O, N, Ne, Mg, S, Si and Fe) and following in detail the stellar population and chemical enrichment by SN-Ia, SN-II, AGB \citep{2003MNRAS.342.1025T,2007MNRAS.382.1050T} and cooling tables from \citet{2009MNRAS.399..574W};
 black holes and associated Active Galactic Nucleai (AGN) feedback \citep{2005MNRAS.361..776S} with various improvements \citep{2010MNRAS.401.1670F,2014MNRAS.442.2304H} for the treatment of the black hole sink particles and the different feedback modes;
 isotropic thermal conduction of 1/20 of standard Spitzer value \citep{2004ApJ...606L..97D};
 low viscosity scheme to track turbulence \citep{2005MNRAS.364..753D,2016MNRAS.455.2110B};
 higher order SPH kernels \citep{2012MNRAS.425.1068D};
 passive magnetic fields \citep{2009MNRAS.398.1678D}.
Halos are identified using \textsc{subfind} \citep{2001MNRAS.328..726S, 2009MNRAS.399..497D}, where the center of a halo is defined as the position of the particle with the minimum of the gravitational potential. The virial mass, $M_\mathrm{vir}$ is defined through the spherical-overdensity as predicted by the generalised spherical top-hat collapse model \citep{1996MNRAS.282..263E} and, in particular, it is referred to $R_{\mathrm vir}$, whose overdensity to the critical density follows Eq.~6 of \citet{Bryan98}, which correspond to $\approx117$ for the given redshift and cosmology. The radii $R_{200,\mathrm m}$ and $R_{500,\mathrm c}$ are defined as a spherical-overdensity of 200 (respectively 500) to the mean (respectively critical) density in the chosen cosmology.  

As shown in previous studies, the galaxy physics implemented in the \textit{Magneticum} simulations leads to successfully reproducing basic galaxy properties like the stellar mass-function \citep{2017ARA&A..55...59N,2022arXiv220109068L}, the environmental impact of galaxy clusters on galaxy properties \citep{2019MNRAS.488.5370L} and the appearance of post-starburst galaxies \citep{2021MNRAS.506.4516L} as well as the associated AGN population at various redshifts \citep{2014MNRAS.442.2304H,2016MNRAS.458.1013S,2018MNRAS.481.2213B}. At cluster scale, the \textit{Magneticum} simulations has demonstrated to reproduce the observable X-ray luminosity-relation \citep{2013MNRAS.428.1395B}, the pressure profile of the ICM \citep{2017MNRAS.469.3069G} and the chemical composition \citep{2017Galax...5...35D,2018SSRv..214..123B} of the ICM, the high concentration observed in fossil groups \citep{2019MNRAS.486.4001R},  as well as the gas properties in between galaxy clusters \citep{2021arXiv210614542B}. On larger scale, the \textit{Magneticum} simulations demonstrated to reproduce the observed SZ-Power spectrum \cite{2016MNRAS.463.1797D} as well as the observed thermal history of the Universe \citep{2021PhRvD.104h3538Y}.        

To build our sample, we divide the mass range between $M_{\rm vir}$, from $10^{13} \ h^{-1}M_{\odot}$ and $>5\cdot10^{14} \ h^{-1}M_{\odot}$ (corresponding to $M_{\rm 500,\mathrm c}$ between $4.2\cdot10^{12} \ h^{-1}M_{\odot}$ and $5.4\cdot10^{14} \ h^{-1}M_{\odot}$ for the less massive and the most massive system, respectively) in 14 equal bins in the logarithmic space.
For each bin, we randomly selected 10 objects from simulated \textit{Box2b}. Therefore, the final sample is composed by a total of 140 galaxy groups and clusters where the smallest halo is represented by $4.6\cdot10^4$ particles in the radial range shown in Fig.~\ref{fig:dens_maps}, which gives the density profiles for each objects, as well as the median values computed in the 14 mass bins. 

In addition, Fig.~\ref{fig:dens_maps} also shows two gas density maps for the most massive and less one object in our sample, produced using the software \textsc{SMAC} \citep{2005MNRAS.363...29D}. 
To have also an upper limit for the radial range of our analysis, we consider boundary of our system the position of the accretion shocks. Different definition of the position of accretion shock could be found in literature \citep{Zhang20,Aung21}. In our work, we assume that the position of the accretion shock is located at same radius at which the gas entropy profile reach the peak \citep{2011MNRAS.418..960V}. Expressing this distance in function of $R_{200,\mathrm m}$, we find that the median distance of accretion shock from system center is $3(\pm1)R_{200,\mathrm m}$. Moreover, for our sample, we find also that $R_{200,\mathrm m}$ is $2.3(\pm0.1)R_{500,\mathrm c}$. Combing these values and wanting to characterize the baryon and gas fraction inside the entire volume of our galaxy groups and clusters, we extend our analysis up to $10R_{500,\mathrm c}$. A more detailed study about the position of accretion shock in our sample will be part of a forthcoming and dedicated paper.

\begin{figure*}[ht]
\includegraphics[width=0.49\textwidth]{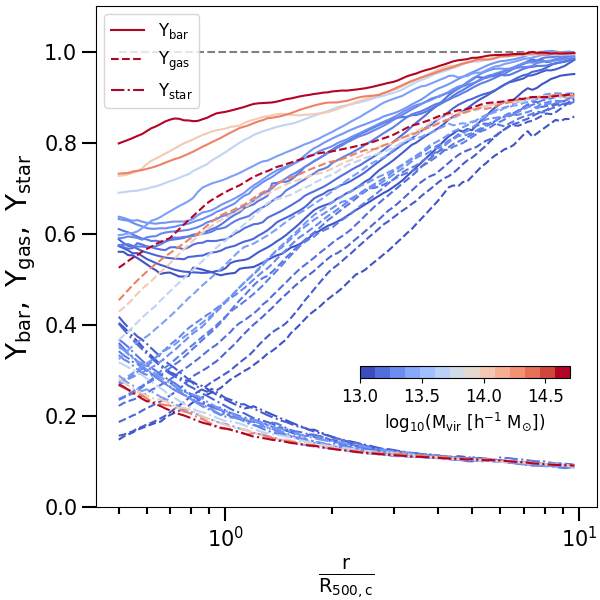}
\includegraphics[width=0.49\textwidth]{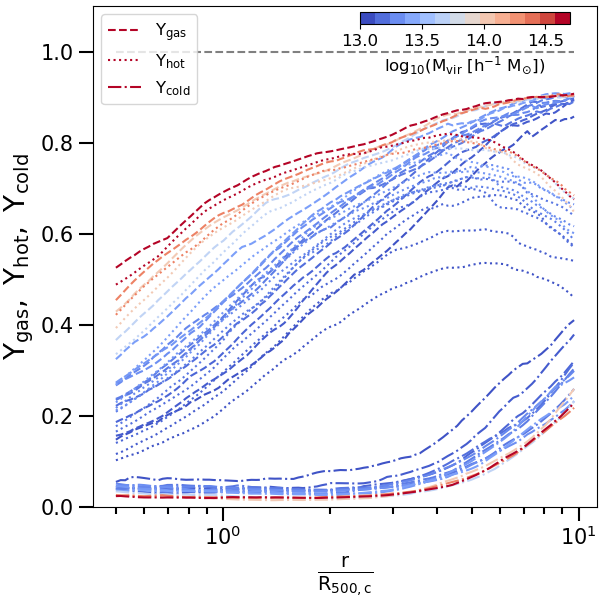}
\caption{(Left) Radial profiles of baryon, gas and star depletion, from $0.5R_{500, \mathrm c}$ up to $10R_{500, \mathrm c}$. The lines represent the median profiles in each mass bin, while the different line-styles represent baryon (solid), gas (dashed) or star depletion (dash-dotted).
(Right) Same as the plot on the left, but representing only three different gas particle selections:  gas particles are in \textbf{dashed}, hot gas with temperature greater than 0.1 keV are in \textbf{dotted}, while cold gas with temperature below 0.1 keV are in dash-dotted.}
\label{fig:ybar_ygas_ystar}
\end{figure*}

\section{The gas and baryon fractions out to $10 R_{500, \mathrm c}$}
\label{sec:results}

We define the baryon, gas and star fraction as:
\begin{align}
  f_{\rm bar}(<r) &= (m_{\rm gas}(<r)+m_{\rm star}(<r)+m_{\rm BH}(<r))/m_{\rm tot}(<r) \\
    f_{\rm gas}(<r) &= m_{\rm gas}(<r)/m_{\rm tot}(<r) \\
    f_{\rm star}(<r) &= m_{\rm star}(<r)/m_{\rm tot}(<r)
\end{align}
where $r$ is the radial distance from the cluster or group center. The different $m_{\rm i}(<r)$ are referred to different particles type (gas, stars or black holes), while $m_{\rm tot}(<r)$ is the sum of the previous masses and the dark matter up to the radial shell $r$. 
Regarding the black holes, which are no longer considered in our analysis, their median mass is $7\cdot10^{8} \ h^{-1} M_{\odot}$, while their contribution to the to the total baryon budget rapidly decreases with the radial distance from the center of the system, from $0.2\%$ to $0.02\%$, without strong dependencies with the halo mass.
In the following we use the depletion parameter $Y$, defined as:
\begin{equation}
    Y(<r) = f(<r)/(\Omega_{\rm b}/\Omega_{\rm m})
\end{equation}
where $f(<r)$ could assume any definition given above and $\Omega_{\rm b}/\Omega_{\rm m}=0.168$, the cosmological value of baryon over total matter adopted for \textit{Magneticum} simulations.

Firstly, we compare our results  with the numerical and observational literature.  In particular, in left panel of Fig.~\ref{fig:ygas_obs_sim}, we compare our gas depletion with the values reported  by \citet{Kravtsov05} and \citet{Planelles13}. 
From left panel of Fig.~\ref{fig:ygas_obs_sim}, we can conclude that the gas depletion factor we recover at $R_{500,c}$ for the most massive systems is in line with the numerical literature. 
However, we expect that on group mass scales, the potential well is less effective in bounding the accreting baryons and balancing the dispersive actions of feedback from AGN and winds from star formation activity. Therefore, we expect a decrease of the gas depletion factor moving towards lower mass scales \citep[see e.g.][]{Eckert21,Akino22}.
In Fig.~\ref{fig:ygas_obs_sim}, we show how our simulated dataset is able to recover the expected and observed behaviour within $R_{500,\mathrm c}$.

In Fig.~\ref{fig:ybar_ygas_ystar}, we show the distributions of the baryon ($Y_{\rm bar}$), gas ($Y_{\rm gas}$) and star ($Y_{\rm star}$) depletion factors as function of the radius (between $0.5R_{500, \mathrm c}$ and $10R_{500, \mathrm c}$) and halo mass. 
We show that, although the baryon fraction always is $\geq 50\%$ of the cosmological value $\Omega_{\rm b}/\Omega_{\rm m}$ in each mass bin, it reaches the cosmological value only at radii larger than $5R_{500, \mathrm c}$ and only in the most massive systems. Indeed, $Y_{\rm bar}$ is greater than $0.99$ at $r>5R_{500, \mathrm c}$ in the sub-sample D (more massive systems in our catalogue), whereas it has a value of about $0.83$ at $5R_{500, \mathrm c}$ and $0.95$ at $10R_{500, \mathrm c}$ (see Tab.~\ref{tab:fixradii}).
Similar trends are observed for the gas depletion factor $Y_{\rm gas}$. Also in this case, the larger is the radii and higher is the depletion factor. 
Both for the baryonic matter and for the hot gas alone, less massive systems show a steeper increase in the depletion factor with the radius. 
On the contrary, the stellar depletion factor $Y_{\rm star}$ decreases with the radius, and is higher in less massive systems.
At larger radii, this behavior is less prominent and both galaxy groups and clusters show constant and similar values of $Y_{\rm star} = 0.09^{+0.01}_{-0.01}$, confirming that the in situ enrichment does not play a significant role, and its uniform metal abundance is rather the consequence of the accretion of pre-enriched (at z > 2) gas \citep[see][]{biffi18}. On the other hand, a more efficient production of stars is required by the larger $Y_{\rm star}$ estimated in less massive haloes within $R_{500, \mathrm c}$.

Focusing on the gas component of our systems, we define $f_{\rm hot}$ and $f_{\rm cold}$:
\begin{align}
    f_{\rm hot}(<r) &= m_{\rm hot}(<r)/m_{\rm tot}(<r) \\
     f_{\rm cold}(<r) &= m_{\rm cold}(<r)/m_{\rm tot}(<r)
\end{align}
meaning a selection of gas particles based on the temperature, $hot$ for particles with temperature greater than 0.1 keV and $cold$ for the others. 
On right panel of Fig.~\ref{fig:ybar_ygas_ystar}, we show the radial behaviour of these quantities compared to $Y_{\rm gas}$. We note that the cold component of gas content became important at very large radii, over $6R_{500, \mathrm c}$. Moreover, this cold component shows larger values for galaxy groups, for which it is larger than 0.2 at radii greater than $8R_{500, \mathrm c}$. 
The hot phase presents a drop for all the mass bins at radii larger than $6R_{500, \mathrm c}$, where we locate the accretion shocks on average in our systems (see Sect.~\ref{sec:methods}). 
This means that at larger radii we are investigating also the contribution from the environment in which galaxy groups and clusters reside, with a decreasing contribution to the expected gas emissivity in X-ray. 

\subsection{The radial trend of gas metallicity} \label{sec:resultsmet}

The injection and evolution of metals by SN-Ia, SN-II and AGB stars in \textit{Magneticum} simulations are modeled following \citet{2003MNRAS.342.1025T,2007MNRAS.382.1050T}.
Although the simulation traces various elements individually, we consider here the total metallicity, i.e. the sum of the elements heavier than helium relative to the hydrogen mass.
Then, the total metallicity at each radial shell $r$, $Z_{\rm tot}(r)$, is the mass-weighted sum of the metallicity of the gas particles $i$ with mass $m_{\rm gas,i}$ which belong to the radial shell $r$:
\begin{equation}
    Z_{\rm tot}(r) = \frac{\sum_{\rm i} Z_{\rm tot,i} \cdot m_{\rm gas,i}}{\sum_{\rm i} m_{\rm gas,i}}.
\end{equation}
The radial shells are defined to include a fixed number of 250 particles, to allow a significant statistical analysis of each of them. 
We normalise these values of metallicity to the solar values proposed by \citet{Asplund09}: $Z_{\odot}=0.0142$.

\begin{figure}
\includegraphics[width=0.49\textwidth]{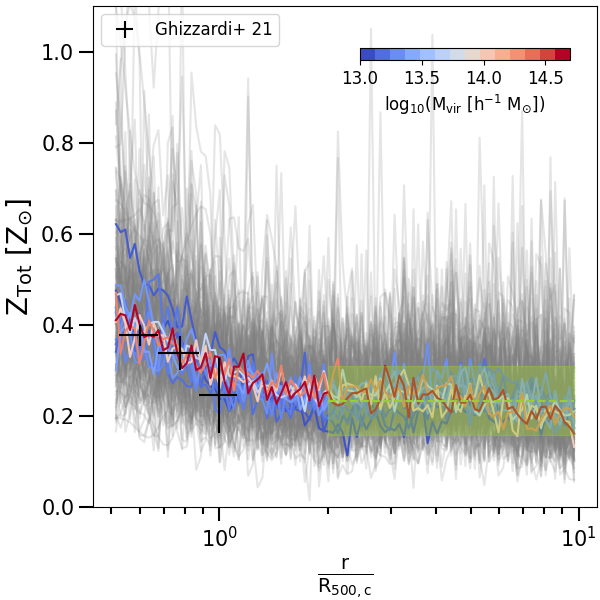}
\caption{Radial profiles of gas metallicity from $0.5R_{500, \mathrm c}$ up to $10R_{500, \mathrm c}$. Gray lines represent single objects, while the colored ones are the mean values computed in mass bins. The green dashed line and its related shadow region show the median, the 16th and 84th percentiles of gas metallicity distribution over $2R_{500, \mathrm c}$ ($0.23^{+0.08}_{-0.08}$). The black dots are the observed estimates of gas metallicity by \citet{Ghizzardi21}.}
\label{fig:metal_gas}
\end{figure}

In Fig.~\ref{fig:metal_gas}, we show the metallicity profiles recovered up to $10 R_{500, \mathrm c}$ in all our halos. 
The profiles are remarkably similar, flattening to a constant value of about $0.23^{+0.08}_{-0.08}$ at $r>2R_{500, \mathrm c}$, and with a negligible dependence upon the halo mass (see median estimates in Tab.~\ref{tab:fixradii}).
Some mass dependency would be expected as consequence of any difference in star-formation activities from group to cluster scales, with the latter being less effective in producing and releasing metals in the environment.
The lack of evidence of such dependency supports the expectations of the early enrichment scenario. \cite{biffi17} \citep[see also][]{biffi18} found that at $r>0.2 R_{180}$ the metallicity is remarkably homogeneous, with almost flat profiles of the elements produced by either SNIa or SNII, mostly as consequence of the widespread displacement of metal-rich gas by early ($z > 2-3$) AGN powerful bursts acting on small high-redshift haloes, and with no significant evolution since redshift $\sim2$. 

\begin{figure*}[ht]
\includegraphics[width=0.315\textwidth]{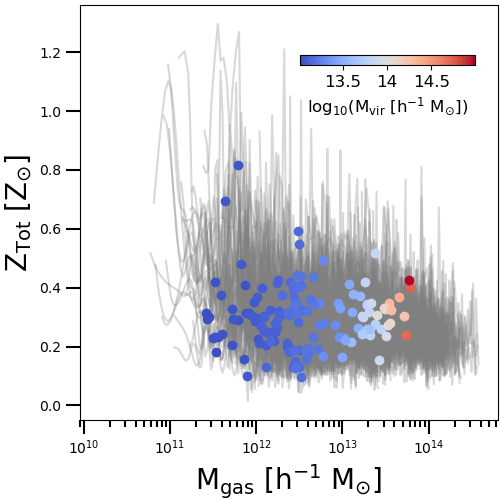}
\includegraphics[width=0.33\textwidth]{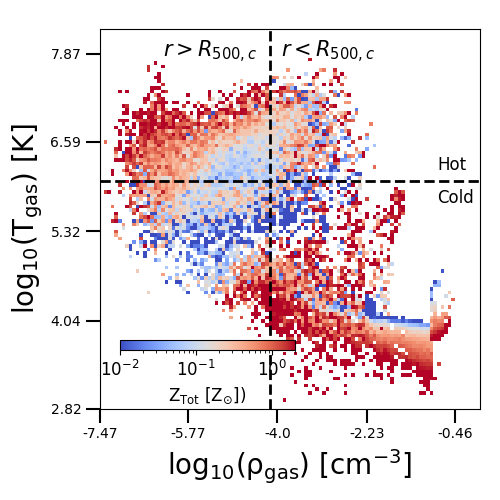}
\includegraphics[width=0.33\textwidth]{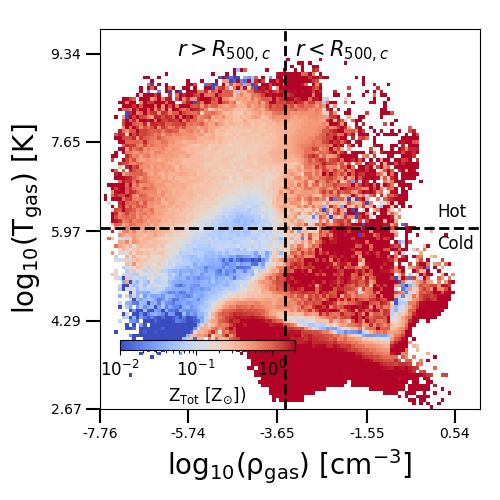}
\caption{(Left) Metallicity as function of gas mass. The grey lines represent the profiles computed for single objects in our sample, whereas the colored dots are the values computed at $R_{500,\mathrm c}$ (again for each groups and clusters present in our sample). The colorcoding is given by the virial mass ($M_{vir}$) of the systems; (Center and Right) Density-temperature phase diagram for the less massive system (center; $M_{vir} = 10^{13} \ h^{-1}M_{\odot}$) and most massive one in our sample (right, $M_{\mathrm vir} = 9.8\cdot10^{14} \ h^{-1}M_{\odot}$). The colording is given by the mass-weighted metallicity in each density-temperature bin. The dashed horizontal black line represent the 0.1keV ($1.16 \cdot 10^6$K) temperature cut, adopted to separate "hot" and "cold" gas phases. The dashed vertical line marks the gas density at $R_{500,\mathrm c}$. On the right of this line there are the particles inside $R_{500,\mathrm c}$, while on the left the particles located at larger radii.}
\label{fig:gasmassvsmet_phasediag}
\end{figure*}

Given the large scatter observed in single radial profiles of metallicity in Fig.~\ref{fig:metal_gas}, we investigate the relations between the masses of the systems and metallicity. From the left panel of Fig.~\ref{fig:gasmassvsmet_phasediag}, where we show the metallicity in function of gas mass, we note how the profiles are still highly scattered, but looking at the values of metallicity computed at $R_{500,\mathrm c}$ (the colored dots in the plot), we also notice that galaxy groups (bluish dots) show an higher spread in metallicity instead of galaxy clusters (reddish dots), with median values of $0.29^{+0.11}_{-0.11}$ and $0.30^{+0.06}_{-0.07}$,respectively. This trend is similar to the findings discussed by \citet{Truong19}. Here the authors focus on the iron contribution in extremely central regions of simulated galaxy clusters and groups ($r<0.1R_{500}$) and they also compare their results with observational work by \citet{Mernier18}. On the other hand, in the central and right panels of Fig.~\ref{fig:gasmassvsmet_phasediag}, we present the density-temperature phase diagram, for the less (center) and most (right) massive objects in our sample, colorcoding by the mass-weighted metallicity of each density-temperature bin. Here, we notice an high spread of metallicity values, both for galaxy group and cluster. Even when we consider the "hot" and "cold" gas phase separately, the scatter in metallicity is still present. From the whole panels of Fig.~\ref{fig:gasmassvsmet_phasediag} we conclude that the observed scatter in radial profiles of metallicity are due to an intrinsic scatter present in the gas particles. Moreover, the coexistence of different gas phases tends to enlarge the observed scatter.

\section{Discussion and Conclusions} \label{sec:conclusions}

While galaxy clusters, in particular the most massive and relaxed ones, are direct proxies of the cosmological baryon fraction \citep[e.g.][]{allen08,ettori09,mantz22}, galaxy groups host an environment where the competing effects of the gravity and the stellar and AGN feedbacks break the self-similar expectations on the gas distribution.
Realistic numerical simulations are thus key to produce accurate three-dimensional models of the likely distribution of gas in and around halos as a function of their mass.

We consider a large sample of 140 galaxy groups and clusters with $M_{vir} > 10^{13} \ h^{-1}M_{\odot}$ at $z=0.25$, simulated at high resolution with the \textit{Magneticum} runs and analysed up to $10R_{500, \mathrm c}$, i.e. roughly beyond the location of the accretion shock.  
We focus our analysis on the radial dependency of the depletion factors $Y_{\rm bar}$, $Y_{\rm gas}$ and $Y_{\rm star}$ (see definitions in Sect.~\ref{sec:results}) well beyond the currently available constraints at $\sim R_{200, \mathrm c}$.
We summarise here our main results:
\begin{itemize}
    \item  We verify that the baryon fraction is always $>$50\% of the cosmological value adopted in our simulations at $r>R_{500, \mathrm c}$ and at any mass bin, and reaches that value at $\sim 5 R_{500, \mathrm c}$, but only in the most massive ones. 
    $Y_{\rm bar}$ in less massive systems, although shows steeper radial profiles, remains 5\% below $\Omega_{\rm b}/\Omega_{\rm m}$ even at $10 R_{500, \mathrm c}$ 
   (see Fig.~\ref{fig:ybar_ygas_ystar} and Tab.~\ref{tab:fixradii}). Similar trends are observed for the gas component. 
    \item Once we divide the gas component in two different phases, {\it hot} when the temperature of the particle is greater than 0.1 keV and {\it cold} for the remaining gas, the contribution given by the cold phase is quite negligible inside $6R_{500,\mathrm c}$, whereas beyond this limit, especially for less massive objects, it gradually becomes more important, reaching 45\% of entire gas contribution for galaxy groups and 25\% for galaxy clusters at $10R_{500,\mathrm c}$ (see Fig.~\ref{fig:ybar_ygas_ystar}). 
    The hot, X-ray detectable, phase is a good tracer of the total amount of gas at all the masses, and represent always more than 90\% (70\%) of it in the most (less) massive objects within $5R_{500,\mathrm c}$.
    At about $6R_{500,\mathrm c}$, $Y_{\rm hot}$ shows a drop for all mass bins. This drop is consistent with the position of accretion shock
    \item We find that both the radial and mass dependency of $Y_{\rm bar}$, $Y_{\rm gas}$ and $Y_{\rm hot}$ (see Fig.~\ref{fig:ybar_ygas_ystar}) are predictable and follow the following functional form 
    \begin{equation} \label{eq:functionalform}
    Y_{i} = \alpha \cdot w^\beta \cdot x^{\gamma + \delta*w}
    \end{equation}
    where $w = M_{500, \mathrm c} / 5\cdot10^{14} \ h^{-1}M_{\odot}$, $x = r /R_{500, \mathrm c}$, while $\alpha$, $\beta$, $\gamma$ and $\delta$ are the free parameters, that are quoted in Tab.~\ref{tab:fitdepletion}. The set of parameters $(\alpha,\beta,\gamma,\delta)= (0.90, 0.12, 0.23, -0.22)$ provides a description within 2\% of $Y_{\rm bar}$ (see Fig.~\ref{fig:ybar_fit}). 
    \item $Y_{\rm star}$ decreases systematically with increasing radii. While the contribution of the stellar mass fraction in the central region ($<2R_{500,\mathrm c}$) is higher in less massive systems, this difference disappear at larger radii, and both galaxy groups and clusters reach a value of $Y_{\rm star} = 0.09^{+0.01}_{-0.01}$. In these regions, the metal distribution becomes constant, with a median value of $0.23^{+0.08}_{-0.08}$ at $r>2R_{500,\mathrm c}$, with no dependency on the halo mass (see Fig.~\ref{fig:metal_gas} and Tab.~\ref{tab:fixradii}).
   These flat mass-independent metallicity profiles support a scenario in which the gas present in massive halos is early enriched, with a negligible contribution from recent star-formation processes (that is indeed expected to be mass-dependent). This result confirms, and extend to much larger radii, what has been obtained from recent work on numerical simulations \citep[see e.g.][]{biffi17,biffi18}.
\end{itemize}

This study has explored the properties of the baryons in massive ($M_{vir} > 10^{13} \ h^{-1} M_{\odot}$) halos well outside the virialized regions, where material from the cosmic field is still accreted.
We have shown convincingly that these halos converge in a different way (depending on their mass) to common properties in terms of baryon, gas, stellar, and metal distribution as they approach the mean location of the accretion shock ($\sim 6 R_{500, \mathrm c}$).
This work allows to make firm predictions on the average distribution of the baryons in their different phases in regions both with current ($r < R_{500, \mathrm c}$) and possible future observational constraints, indicating, for instance, where to look for the gas that is not (X-ray) detected on the mass scales of the galaxy groups.

\section*{Acknowledgements}
S.E. acknowledges financial contribution from the contracts ASI-INAF Athena 2019-27-HH.0,
``Attivit\`a di Studio per la comunit\`a scientifica di Astrofisica delle Alte Energie e Fisica Astroparticellare''
(Accordo Attuativo ASI-INAF n. 2017-14-H.0), INAF mainstream project 1.05.01.86.10, and
from the European Union’s Horizon 2020 Programme under the AHEAD2020 project (grant agreement n. 871158). M.A. and F.V. acknowledge the financial support by the H2020 initiative, through the ERC StG MAGCOW (n. 714196).  A.R. acknowledges support from the grant PRIN-MIUR 2017 WSCC32.
KD acknowledges supported by the Excellence Cluster ORIGINS which is funded by the Deutsche Forschungsgemeinschaft (DFG, German Research Foundation) under Germany´s Excellence Strategy – EXC-2094 – 390783311 and funding for the COMPLEX project from the European Research Council (ERC) under the European Union’s Horizon 2020 research and innovation program grant agreement ERC-2019-AdG 882679. The calculations for the hydro-dynamical simulations were carried out at the Leibniz Supercomputer Center (LRZ) under the project pr83li. We are especially grateful for the support by M. Petkova through the Computational Center for Particle and Astrophysics (C2PAP). 
\bibliographystyle{aa}
\bibliography{biblio}

\begin{thebibliography}{70}
\expandafter\ifx\csname natexlab\endcsname\relax\def\natexlab#1{#1}\fi

\bibitem[{{Akino} {et~al.}(2022){Akino}, {Eckert}, {Okabe}, {Sereno}, {Umetsu},
  {Oguri}, {Gastaldello}, {Chiu}, {Ettori}, {Evrard}, {Farahi}, {Maughan},
  {Pierre}, {Ricci}, {Valtchanov}, {McCarthy}, {McGee}, {Miyazaki},
  {Nishizawa}, \& {Tanaka}}]{Akino22}
{Akino}, D., {Eckert}, D., {Okabe}, N., {et~al.} 2022, \pasj, 74, 175

\bibitem[{{Allen} {et~al.}(2011){Allen}, {Evrard}, \& {Mantz}}]{allen11}
{Allen}, S.~W., {Evrard}, A.~E., \& {Mantz}, A.~B. 2011, \araa, 49, 409

\bibitem[{{Allen} {et~al.}(2008){Allen}, {Rapetti}, {Schmidt}, {Ebeling},
  {Morris}, \& {Fabian}}]{allen08}
{Allen}, S.~W., {Rapetti}, D.~A., {Schmidt}, R.~W., {et~al.} 2008, \mnras, 383,
  879

\bibitem[{{Angelinelli} {et~al.}(2020){Angelinelli}, {Vazza}, {Giocoli},
  {Ettori}, {Jones}, {Brunetti}, {Br{\"u}ggen}, \& {Eckert}}]{Angelinelli20}
{Angelinelli}, M., {Vazza}, F., {Giocoli}, C., {et~al.} 2020, \mnras, 495, 864

\bibitem[{{Asplund} {et~al.}(2009){Asplund}, {Grevesse}, {Sauval}, \&
  {Scott}}]{Asplund09}
{Asplund}, M., {Grevesse}, N., {Sauval}, A.~J., \& {Scott}, P. 2009, \araa, 47,
  481

\bibitem[{{Aung} {et~al.}(2021){Aung}, {Nagai}, \& {Lau}}]{Aung21}
{Aung}, H., {Nagai}, D., \& {Lau}, E.~T. 2021, \mnras, 508, 2071

\bibitem[{{Beck} {et~al.}(2016){Beck}, {Murante}, {Arth}, {Remus}, {Teklu},
  {Donnert}, {Planelles}, {Beck}, {F{\"o}rster}, {Imgrund}, {Dolag}, \&
  {Borgani}}]{2016MNRAS.455.2110B}
{Beck}, A.~M., {Murante}, G., {Arth}, A., {et~al.} 2016, \mnras, 455, 2110

\bibitem[{{Bertschinger}(1985)}]{Bertschinger85}
{Bertschinger}, E. 1985, \apjs, 58, 39

\bibitem[{{Biffi} {et~al.}(2013){Biffi}, {Dolag}, \&
  {B{\"o}hringer}}]{2013MNRAS.428.1395B}
{Biffi}, V., {Dolag}, K., \& {B{\"o}hringer}, H. 2013, \mnras, 428, 1395

\bibitem[{{Biffi} {et~al.}(2018{\natexlab{a}}){Biffi}, {Dolag}, \&
  {Merloni}}]{2018MNRAS.481.2213B}
{Biffi}, V., {Dolag}, K., \& {Merloni}, A. 2018{\natexlab{a}}, \mnras, 481,
  2213

\bibitem[{{Biffi} {et~al.}(2021){Biffi}, {Dolag}, {Reiprich}, {Veronica},
  {Ramos-Ceja}, {Bulbul}, {Ota}, \& {Ghirardini}}]{2021arXiv210614542B}
{Biffi}, V., {Dolag}, K., {Reiprich}, T.~H., {et~al.} 2021, arXiv e-prints,
  arXiv:2106.14542

\bibitem[{{Biffi} {et~al.}(2018{\natexlab{b}}){Biffi}, {Mernier}, \&
  {Medvedev}}]{2018SSRv..214..123B}
{Biffi}, V., {Mernier}, F., \& {Medvedev}, P. 2018{\natexlab{b}}, \ssr, 214,
  123

\bibitem[{{Biffi} {et~al.}(2017){Biffi}, {Planelles}, {Borgani}, {Fabjan},
  {Rasia}, {Murante}, {Tornatore}, {Dolag}, {Granato}, {Gaspari}, \&
  {Beck}}]{biffi17}
{Biffi}, V., {Planelles}, S., {Borgani}, S., {et~al.} 2017, \mnras, 468, 531

\bibitem[{{Biffi} {et~al.}(2018{\natexlab{c}}){Biffi}, {Planelles}, {Borgani},
  {Rasia}, {Murante}, {Fabjan}, \& {Gaspari}}]{biffi18}
{Biffi}, V., {Planelles}, S., {Borgani}, S., {et~al.} 2018{\natexlab{c}},
  \mnras, 476, 2689

\bibitem[{{Bryan} \& {Norman}(1998)}]{Bryan98}
{Bryan}, G.~L. \& {Norman}, M.~L. 1998, \apj, 495, 80

\bibitem[{{Dehnen} \& {Aly}(2012)}]{2012MNRAS.425.1068D}
{Dehnen}, W. \& {Aly}, H. 2012, \mnras, 425, 1068

\bibitem[{{Dolag} {et~al.}(2009){Dolag}, {Borgani}, {Murante}, \&
  {Springel}}]{2009MNRAS.399..497D}
{Dolag}, K., {Borgani}, S., {Murante}, G., \& {Springel}, V. 2009, \mnras, 399,
  497

\bibitem[{{Dolag} {et~al.}(2005{\natexlab{a}}){Dolag}, {Hansen}, {Roncarelli},
  \& {Moscardini}}]{2005MNRAS.363...29D}
{Dolag}, K., {Hansen}, F.~K., {Roncarelli}, M., \& {Moscardini}, L.
  2005{\natexlab{a}}, \mnras, 363, 29

\bibitem[{{Dolag} {et~al.}(2004){Dolag}, {Jubelgas}, {Springel}, {Borgani}, \&
  {Rasia}}]{2004ApJ...606L..97D}
{Dolag}, K., {Jubelgas}, M., {Springel}, V., {Borgani}, S., \& {Rasia}, E.
  2004, \apjl, 606, L97

\bibitem[{{Dolag} {et~al.}(2016){Dolag}, {Komatsu}, \&
  {Sunyaev}}]{2016MNRAS.463.1797D}
{Dolag}, K., {Komatsu}, E., \& {Sunyaev}, R. 2016, \mnras, 463, 1797

\bibitem[{{Dolag} {et~al.}(2017){Dolag}, {Mevius}, \&
  {Remus}}]{2017Galax...5...35D}
{Dolag}, K., {Mevius}, E., \& {Remus}, R.-S. 2017, Galaxies, 5, 35

\bibitem[{{Dolag} \& {Stasyszyn}(2009)}]{2009MNRAS.398.1678D}
{Dolag}, K. \& {Stasyszyn}, F. 2009, \mnras, 398, 1678

\bibitem[{{Dolag} {et~al.}(2005{\natexlab{b}}){Dolag}, {Vazza}, {Brunetti}, \&
  {Tormen}}]{2005MNRAS.364..753D}
{Dolag}, K., {Vazza}, F., {Brunetti}, G., \& {Tormen}, G. 2005{\natexlab{b}},
  \mnras, 364, 753

\bibitem[{{Eckert} {et~al.}(2016){Eckert}, {Ettori}, {Coupon}, {Gastaldello},
  {Pierre}, {Melin}, {Le Brun}, {McCarthy}, {Adami}, {Chiappetti}, {Faccioli},
  {Giles}, {Lavoie}, {Lef{\`e}vre}, {Lieu}, {Mantz}, {Maughan}, {McGee},
  {Pacaud}, {Paltani}, {Sadibekova}, {Smith}, \& {Ziparo}}]{Eckert16}
{Eckert}, D., {Ettori}, S., {Coupon}, J., {et~al.} 2016, \aap, 592, A12

\bibitem[{{Eckert} {et~al.}(2021){Eckert}, {Gaspari}, {Gastaldello}, {Le Brun},
  \& {O'Sullivan}}]{Eckert21}
{Eckert}, D., {Gaspari}, M., {Gastaldello}, F., {Le Brun}, A. M.~C., \&
  {O'Sullivan}, E. 2021, Universe, 7, 142

\bibitem[{{Eke} {et~al.}(1996){Eke}, {Cole}, \& {Frenk}}]{1996MNRAS.282..263E}
{Eke}, V.~R., {Cole}, S., \& {Frenk}, C.~S. 1996, \mnras, 282, 263

\bibitem[{{Ettori}(2015)}]{Ettori15}
{Ettori}, S. 2015, \mnras, 446, 2629

\bibitem[{{Ettori} {et~al.}(2006){Ettori}, {Dolag}, {Borgani}, \&
  {Murante}}]{Ettori06}
{Ettori}, S., {Dolag}, K., {Borgani}, S., \& {Murante}, G. 2006, \mnras, 365,
  1021

\bibitem[{{Ettori} {et~al.}(2009){Ettori}, {Morandi}, {Tozzi}, {Balestra},
  {Borgani}, {Rosati}, {Lovisari}, \& {Terenziani}}]{ettori09}
{Ettori}, S., {Morandi}, A., {Tozzi}, P., {et~al.} 2009, \aap, 501, 61

\bibitem[{{Fabjan} {et~al.}(2010){Fabjan}, {Borgani}, {Tornatore}, {Saro},
  {Murante}, \& {Dolag}}]{2010MNRAS.401.1670F}
{Fabjan}, D., {Borgani}, S., {Tornatore}, L., {et~al.} 2010, \mnras, 401, 1670

\bibitem[{{Gal{\'a}rraga-Espinosa} {et~al.}(2021){Gal{\'a}rraga-Espinosa},
  {Langer}, \& {Aghanim}}]{Galarraga-Espinosa21}
{Gal{\'a}rraga-Espinosa}, D., {Langer}, M., \& {Aghanim}, N. 2021, arXiv
  e-prints, arXiv:2109.06198

\bibitem[{{Gastaldello} {et~al.}(2021){Gastaldello}, {Simionescu}, {Mernier},
  {Biffi}, {Gaspari}, {Sato}, \& {Matsushita}}]{Gastaldello21}
{Gastaldello}, F., {Simionescu}, A., {Mernier}, F., {et~al.} 2021, Universe, 7,
  208

\bibitem[{{Ghirardini} {et~al.}(2019){Ghirardini}, {Eckert}, {Ettori},
  {Pointecouteau}, {Molendi}, {Gaspari}, {Rossetti}, {De Grandi}, {Roncarelli},
  {Bourdin}, {Mazzotta}, {Rasia}, \& {Vazza}}]{Ghirardini:2019}
{Ghirardini}, V., {Eckert}, D., {Ettori}, S., {et~al.} 2019, \aap, 621, A41

\bibitem[{{Ghizzardi} {et~al.}(2021){Ghizzardi}, {Molendi}, {van der Burg}, {De
  Grandi}, {Bartalucci}, {Gastaldello}, {Rossetti}, {Biffi}, {Borgani},
  {Eckert}, {Ettori}, {Gaspari}, {Ghirardini}, \& {Rasia}}]{Ghizzardi21}
{Ghizzardi}, S., {Molendi}, S., {van der Burg}, R., {et~al.} 2021, \aap, 646,
  A92

\bibitem[{{Gunn} \& {Gott}(1972)}]{Gunn72}
{Gunn}, J.~E. \& {Gott}, J.~Richard, I. 1972, \apj, 176, 1

\bibitem[{{Gupta} {et~al.}(2017){Gupta}, {Saro}, {Mohr}, {Dolag}, \&
  {Liu}}]{2017MNRAS.469.3069G}
{Gupta}, N., {Saro}, A., {Mohr}, J.~J., {Dolag}, K., \& {Liu}, J. 2017, \mnras,
  469, 3069

\bibitem[{{Haider} {et~al.}(2016){Haider}, {Steinhauser}, {Vogelsberger},
  {Genel}, {Springel}, {Torrey}, \& {Hernquist}}]{Haider16}
{Haider}, M., {Steinhauser}, D., {Vogelsberger}, M., {et~al.} 2016, \mnras,
  457, 3024

\bibitem[{{Hirschmann} {et~al.}(2014){Hirschmann}, {Dolag}, {Saro}, {Bachmann},
  {Borgani}, \& {Burkert}}]{2014MNRAS.442.2304H}
{Hirschmann}, M., {Dolag}, K., {Saro}, A., {et~al.} 2014, \mnras, 442, 2304

\bibitem[{{Komatsu} {et~al.}(2011){Komatsu}, {Smith}, {Dunkley}, {Bennett},
  {Gold}, {Hinshaw}, {Jarosik}, {Larson}, {Nolta}, {Page}, {Spergel},
  {Halpern}, {Hill}, {Kogut}, {Limon}, {Meyer}, {Odegard}, {Tucker}, {Weiland},
  {Wollack}, \& {Wright}}]{2011ApJS..192...18K}
{Komatsu}, E., {Smith}, K.~M., {Dunkley}, J., {et~al.} 2011, \apjs, 192, 18

\bibitem[{{Kravtsov} {et~al.}(2005){Kravtsov}, {Nagai}, \&
  {Vikhlinin}}]{Kravtsov05}
{Kravtsov}, A.~V., {Nagai}, D., \& {Vikhlinin}, A.~A. 2005, \apj, 625, 588

\bibitem[{{Limber}(1959)}]{Limber59}
{Limber}, D.~N. 1959, \apj, 130, 414

\bibitem[{{Lotz} {et~al.}(2021){Lotz}, {Dolag}, {Remus}, \&
  {Burkert}}]{2021MNRAS.506.4516L}
{Lotz}, M., {Dolag}, K., {Remus}, R.-S., \& {Burkert}, A. 2021, \mnras, 506,
  4516

\bibitem[{{Lotz} {et~al.}(2019){Lotz}, {Remus}, {Dolag}, {Biviano}, \&
  {Burkert}}]{2019MNRAS.488.5370L}
{Lotz}, M., {Remus}, R.-S., {Dolag}, K., {Biviano}, A., \& {Burkert}, A. 2019,
  \mnras, 488, 5370

\bibitem[{{Lovisari} {et~al.}(2015){Lovisari}, {Reiprich}, \&
  {Schellenberger}}]{Lovisari15}
{Lovisari}, L., {Reiprich}, T.~H., \& {Schellenberger}, G. 2015, \aap, 573,
  A118

\bibitem[{{Lustig} {et~al.}(2022){Lustig}, {Strazzullo}, {Remus}, {D'Eugenio},
  {Daddi}, {Burkert}, {De Lucia}, {Delvecchio}, {Dolag}, {Fontanot}, {Gobat},
  {Mohr}, {Onodera}, {Pannella}, {Pillepich}, \&
  {Renzini}}]{2022arXiv220109068L}
{Lustig}, P., {Strazzullo}, V., {Remus}, R.-S., {et~al.} 2022, arXiv e-prints,
  arXiv:2201.09068

\bibitem[{{Mantz} {et~al.}(2022){Mantz}, {Morris}, {Allen}, {Canning},
  {Baumont}, {Benson}, {Bleem}, {Ehlert}, {Floyd}, {Herbonnet}, {Kelly},
  {Liang}, {von der Linden}, {McDonald}, {Rapetti}, {Schmidt}, {Werner}, \&
  {Wright}}]{mantz22}
{Mantz}, A.~B., {Morris}, R.~G., {Allen}, S.~W., {et~al.} 2022, \mnras, 510,
  131

\bibitem[{{McCarthy} {et~al.}(2017){McCarthy}, {Schaye}, {Bird}, \& {Le
  Brun}}]{McCarthy17}
{McCarthy}, I.~G., {Schaye}, J., {Bird}, S., \& {Le Brun}, A. M.~C. 2017,
  \mnras, 465, 2936

\bibitem[{{Mernier} {et~al.}(2018){Mernier}, {de Plaa}, {Werner}, {Kaastra},
  {Raassen}, {Gu}, {Mao}, {Urdampilleta}, {Truong}, \&
  {Simionescu}}]{Mernier18}
{Mernier}, F., {de Plaa}, J., {Werner}, N., {et~al.} 2018, \mnras, 478, L116

\bibitem[{{Naab} \& {Ostriker}(2017)}]{2017ARA&A..55...59N}
{Naab}, T. \& {Ostriker}, J.~P. 2017, \araa, 55, 59

\bibitem[{{Nugent} {et~al.}(2020){Nugent}, {Dai}, \& {Sun}}]{Nugent20}
{Nugent}, J.~M., {Dai}, X., \& {Sun}, M. 2020, \apj, 899, 160

\bibitem[{{Oppenheimer} {et~al.}(2021){Oppenheimer}, {Babul}, {Bah{\'e}},
  {Butsky}, \& {McCarthy}}]{Oppenheimer21}
{Oppenheimer}, B.~D., {Babul}, A., {Bah{\'e}}, Y., {Butsky}, I.~S., \&
  {McCarthy}, I.~G. 2021, Universe, 7, 209

\bibitem[{{Pillepich} {et~al.}(2018){Pillepich}, {Nelson}, {Hernquist},
  {Springel}, {Pakmor}, {Torrey}, {Weinberger}, {Genel}, {Naiman}, {Marinacci},
  \& {Vogelsberger}}]{Pillepich18}
{Pillepich}, A., {Nelson}, D., {Hernquist}, L., {et~al.} 2018, \mnras, 475, 648

\bibitem[{{Planelles} {et~al.}(2013){Planelles}, {Borgani}, {Dolag}, {Ettori},
  {Fabjan}, {Murante}, \& {Tornatore}}]{Planelles13}
{Planelles}, S., {Borgani}, S., {Dolag}, K., {et~al.} 2013, \mnras, 431, 1487

\bibitem[{{Ragagnin} {et~al.}(2019){Ragagnin}, {Dolag}, {Moscardini},
  {Biviano}, \& {D'Onofrio}}]{2019MNRAS.486.4001R}
{Ragagnin}, A., {Dolag}, K., {Moscardini}, L., {Biviano}, A., \& {D'Onofrio},
  M. 2019, \mnras, 486, 4001

\bibitem[{{Roncarelli} {et~al.}(2018){Roncarelli}, {Gaspari}, {Ettori},
  {Biffi}, {Brighenti}, {Bulbul}, {Clerc}, {Cucchetti}, {Pointecouteau}, \&
  {Rasia}}]{2018A&A...618A..39R}
{Roncarelli}, M., {Gaspari}, M., {Ettori}, S., {et~al.} 2018, \aap, 618, A39

\bibitem[{{Springel}(2005)}]{sp05}
{Springel}, V. 2005, \mnras, 364, 1105

\bibitem[{{Springel} {et~al.}(2005){Springel}, {Di Matteo}, \&
  {Hernquist}}]{2005MNRAS.361..776S}
{Springel}, V., {Di Matteo}, T., \& {Hernquist}, L. 2005, \mnras, 361, 776

\bibitem[{{Springel} \& {Hernquist}(2002)}]{2002MNRAS.333..649S}
{Springel}, V. \& {Hernquist}, L. 2002, \mnras, 333, 649

\bibitem[{{Springel} {et~al.}(2001){Springel}, {White}, {Tormen}, \&
  {Kauffmann}}]{2001MNRAS.328..726S}
{Springel}, V., {White}, S. D.~M., {Tormen}, G., \& {Kauffmann}, G. 2001,
  \mnras, 328, 726

\bibitem[{{Steinborn} {et~al.}(2016){Steinborn}, {Dolag}, {Comerford},
  {Hirschmann}, {Remus}, \& {Teklu}}]{2016MNRAS.458.1013S}
{Steinborn}, L.~K., {Dolag}, K., {Comerford}, J.~M., {et~al.} 2016, \mnras,
  458, 1013

\bibitem[{{Sun} {et~al.}(2009){Sun}, {Voit}, {Donahue}, {Jones}, {Forman}, \&
  {Vikhlinin}}]{Sun09}
{Sun}, M., {Voit}, G.~M., {Donahue}, M., {et~al.} 2009, \apj, 693, 1142

\bibitem[{{Tornatore} {et~al.}(2007){Tornatore}, {Borgani}, {Dolag}, \&
  {Matteucci}}]{2007MNRAS.382.1050T}
{Tornatore}, L., {Borgani}, S., {Dolag}, K., \& {Matteucci}, F. 2007, \mnras,
  382, 1050

\bibitem[{{Tornatore} {et~al.}(2003){Tornatore}, {Borgani}, {Springel},
  {Matteucci}, {Menci}, \& {Murante}}]{2003MNRAS.342.1025T}
{Tornatore}, L., {Borgani}, S., {Springel}, V., {et~al.} 2003, \mnras, 342,
  1025

\bibitem[{{Truong} {et~al.}(2019){Truong}, {Rasia}, {Biffi}, {Mernier},
  {Werner}, {Gaspari}, {Borgani}, {Planelles}, {Fabjan}, \&
  {Murante}}]{Truong19}
{Truong}, N., {Rasia}, E., {Biffi}, V., {et~al.} 2019, \mnras, 484, 2896

\bibitem[{{Vall{\'e}s-P{\'e}rez} {et~al.}(2020){Vall{\'e}s-P{\'e}rez},
  {Planelles}, \& {Quilis}}]{VallesPerez20}
{Vall{\'e}s-P{\'e}rez}, D., {Planelles}, S., \& {Quilis}, V. 2020, \mnras, 499,
  2303

\bibitem[{{Vazza} {et~al.}(2011){Vazza}, {Dolag}, {Ryu}, {Brunetti}, {Gheller},
  {Kang}, \& {Pfrommer}}]{2011MNRAS.418..960V}
{Vazza}, F., {Dolag}, K., {Ryu}, D., {et~al.} 2011, \mnras, 418, 960

\bibitem[{{Voit}(2005)}]{Voit2005}
{Voit}, G.~M. 2005, Reviews of Modern Physics, 77, 207

\bibitem[{{Wiersma} {et~al.}(2009){Wiersma}, {Schaye}, {Theuns}, {Dalla
  Vecchia}, \& {Tornatore}}]{2009MNRAS.399..574W}
{Wiersma}, R. P.~C., {Schaye}, J., {Theuns}, T., {Dalla Vecchia}, C., \&
  {Tornatore}, L. 2009, \mnras, 399, 574

\bibitem[{{Young} {et~al.}(2021){Young}, {Komatsu}, \&
  {Dolag}}]{2021PhRvD.104h3538Y}
{Young}, S., {Komatsu}, E., \& {Dolag}, K. 2021, \prd, 104, 083538

\bibitem[{{Zhang} {et~al.}(2020){Zhang}, {Churazov}, {Dolag}, {Forman}, \&
  {Zhuravleva}}]{Zhang20}
{Zhang}, C., {Churazov}, E., {Dolag}, K., {Forman}, W.~R., \& {Zhuravleva}, I.
  2020, \mnras, 494, 4539

\end{thebibliography}

\appendix
\section{Estimates and fitting parameters of the depletion factors}

We provide here details on the values of the depletion parameters under investigations in different mass bins and at various radii (see Tab.~\ref{tab:fixradii}).

We also provide with the best-fit parameters obtained from the fitting function in Eq.~\ref{eq:functionalform} (see Tab.~\ref{tab:fitdepletion}).
Being our findings potentially useful to make predictions and comparisons with observational work, we confine our fitting produce in a radial range similar to one available for present and near future X-rays observations and limited to the central regions of our analysis, from $0.5R_{500,\mathrm c}$ to $2.5R_{500,\mathrm c}$.
The functional form is able to well reproduce the behavior of $Y_{\rm bar}$ at all considered masses, as also shown by the $\Tilde{\chi}^2$ and values of the the median and the maximum deviation of the model from the data ($\Tilde{e}$ and $e_{max}$, respectively) quoted in Tab.~\ref{tab:fitdepletion}. However, for the $Y_{\rm gas}$ and $Y_{\rm hot}$, we note that our fitting procedure gives less strong results than for $Y_{\rm bar}$ case. We use the dispersion around the mean profile as weight to evaluate $\chi^2$.

\begin{table}
\begin{tabular}{c|c|cccc|}
\multicolumn{1}{c}{}&&$R_{500,\mathrm c}$&$3R_{500,\mathrm c}$&$5R_{500,\mathrm c}$&$10R_{500,\mathrm c}$ \\ \hline \hline
\multirow{6}{*}{A}&$Y_{\rm bar}$&$0.51^{+0.09}_{-0.06}$&$0.70^{+0.06}_{-0.05}$&$0.83^{+0.05}_{-0.01}$&$0.95^{+0.04}_{-0.02}$ \\
&$Y_{\rm gas}$&$0.27^{+0.06}_{-0.03}$&$0.57^{+0.05}_{-0.03}$&$0.73^{+0.05}_{-0.03}$&$0.86^{+0.05}_{-0.03}$ \\
&$Y_{\rm hot}$&$0.22^{+0.03}_{-0.02}$&$0.48^{+0.05}_{-0.07}$&$0.53^{+0.06}_{-0.08}$&$0.46^{+0.14}_{-0.10}$ \\
&$Y_{\rm cold}$&$0.06^{+0.03}_{-0.02}$&$0.10^{+0.01}_{-0.02}$&$0.21^{+0.04}_{-0.06}$&$0.41^{+0.08}_{-0.10}$ \\
&$Y_{\rm star}$&$0.24^{+0.03}_{-0.03}$&$0.12^{+0.01}_{-0.01}$&$0.11^{+0.01}_{-0.01}$&$0.09^{+0.01}_{-0.01}$ \\
&$Z_{\rm tot}$&$0.24^{+0.07}_{-0.02}$&$0.17^{+0.05}_{-0.06}$&$0.21^{+0.21}_{-0.05}$&$0.22^{+0.12}_{-0.03}$ \\ \hline
\multirow{6}{*}{B}&$Y_{\rm bar}$&$0.63^{+0.08}_{-0.04}$&$0.85^{+0.03}_{-0.05}$&$0.93^{+0.01}_{-0.02}$&$0.98^{+0.01}_{-0.02}$ \\
&$Y_{\rm gas}$&$0.42^{+0.07}_{-0.05}$&$0.72^{+0.04}_{-0.05}$&$0.82^{+0.02}_{-0.02}$&$0.89^{+0.01}_{-0.03}$ \\
&$Y_{\rm hot}$&$0.37^{+0.03}_{-0.03}$&$0.67^{+0.02}_{-0.03}$&$0.73^{+0.03}_{-0.03}$&$0.64^{+0.08}_{-0.05}$ \\
&$Y_{\rm cold}$&$0.04^{+0.06}_{-0.02}$&$0.04^{+0.02}_{-0.01}$&$0.09^{+0.02}_{-0.03}$&$0.26^{+0.05}_{-0.11}$ \\
&$Y_{\rm star}$&$0.21^{+0.04}_{-0.01}$&$0.12^{+0.01}_{-0.01}$&$0.10^{+0.01}_{-0.01}$&$0.09^{+0.01}_{-0.01}$ \\
&$Z_{\rm tot}$&$0.29^{+0.09}_{-0.06}$&$0.25^{+0.05}_{-0.06}$&$0.26^{+0.04}_{-0.02}$&$0.20^{+0.04}_{-0.05}$ \\ \hline
\multirow{6}{*}{C}&$Y_{\rm bar}$&$0.72^{+0.04}_{-0.07}$&$0.90^{+0.02}_{-0.03}$&$0.95^{+0.03}_{-0.02}$&$0.99^{+0.01}_{-0.01}$ \\
&$Y_{\rm gas}$&$0.53^{+0.03}_{-0.07}$&$0.78^{+0.01}_{-0.03}$&$0.85^{+0.03}_{-0.02}$&$0.90^{+0.01}_{-0.01}$ \\
&$Y_{\rm hot}$&$0.50^{+0.03}_{-0.06}$&$0.74^{+0.01}_{-0.03}$&$0.76^{+0.04}_{-0.03}$&$0.65^{+0.08}_{-0.06}$ \\
&$Y_{\rm cold}$&$0.03^{+0.01}_{-0.01}$&$0.04^{+0.02}_{-0.01}$&$0.08^{+0.03}_{-0.02}$&$0.24^{+0.07}_{-0.07}$ \\
&$Y_{\rm star}$&$0.18^{+0.02}_{-0.01}$&$0.11^{+0.01}_{-0.01}$&$0.10^{+0.01}_{-0.01}$&$0.09^{+0.01}_{-0.01}$ \\
&$Z_{\rm tot}$&$0.28^{+0.07}_{-0.05}$&$0.27^{+0.06}_{-0.05}$&$0.25^{+0.04}_{-0.02}$&$0.21^{+0.08}_{-0.07}$ \\ \hline
\multirow{6}{*}{D}&$Y_{\rm bar}$&$0.87^{+0.03}_{-0.04}$&$0.93^{+0.01}_{-0.02}$&$0.99^{+0.01}_{-0.01}$&$1.00^{+0.01}_{-0.01}$ \\
&$Y_{\rm gas}$&$0.69^{+0.04}_{-0.04}$&$0.82^{+0.02}_{-0.02}$&$0.88^{+0.01}_{-0.01}$&$0.91^{+0.01}_{-0.01}$ \\
&$Y_{\rm hot}$&$0.67^{+0.04}_{-0.05}$&$0.79^{+0.01}_{-0.02}$&$0.81^{+0.01}_{-0.01}$&$0.68^{+0.03}_{-0.02}$ \\
&$Y_{\rm cold}$&$0.02^{+0.01}_{-0.01}$&$0.03^{+0.01}_{-0.01}$&$0.07^{+0.01}_{-0.01}$&$0.23^{+0.02}_{-0.03}$ \\
&$Y_{\rm star}$&$0.17^{+0.02}_{-0.02}$&$0.11^{+0.01}_{-0.01}$&$0.11^{+0.01}_{-0.01}$&$0.09^{+0.01}_{-0.01}$ \\
&$Z_{\rm tot}$&$0.33^{+0.06}_{-0.04}$&$0.28^{+0.04}_{-0.08}$&$0.25^{+0.11}_{-0.08}$&$0.16^{+0.08}_{-0.04}$ \\ \hline
\multicolumn{5}{c}{}\\
\end{tabular}
\caption{Baryons ($Y_{\rm bar}$), gas ($Y_{\rm gas}$), hot gas phase ($Y_{\rm hot}$), cold gas phase ($Y_{\rm cold}$) and stellar ($Y_{\rm star}$) depletion factors and gas metallicity ($Z_{tot}$) computed at four different radii ($1, 3, 5$ and $10$ times $R_{500,\mathrm c}$) for four virial mass bins ($A: \ 10^{13}<M_{vir}/h^{-1}M_{\odot}<2\cdot10^{13} \ (4.2\cdot10^{12}<M_{500,\mathrm c}/h^{-1}M_{\odot}<1.2\cdot10^{13})$; $B: \ 5\cdot10^{13}<M_{vir}/h^{-1}M_{\odot}<6\cdot10^{13} \ (1.6\cdot10^{13}<M_{500,\mathrm c}/h^{-1}M_{\odot}<3.6\cdot10^{13})$; $C: \ 10^{14}<M_{vir}/h^{-1}M_{\odot}<2\cdot10^{14} \ (6.3\cdot10^{13}<M_{500,\mathrm c}/h^{-1}M_{\odot}<10^{14})$; $D: \ M_{vir}/h^{-1}M_{\odot}>5\cdot10^{14} \ (M_{500,\mathrm c}/h^{-1}M_{\odot}>2.4\cdot10^{14})$). Errors are given as 16th and 84th distributions percentiles. Note that the depletion factors are computed within the given radii, differently from the metallicity values which are given within a spherical shell (considering the same reference radii).}
\label{tab:fixradii}
\end{table}

\begin{table}[hb]
\hspace{5pt}
\begin{tabular}{c|ccc|}
&$Y_{\rm bar}$&$Y_{\rm gas}$&$Y_{\rm hot}$ \\ \hline 
$\alpha$&$0.900\pm0.003$&$0.733\pm0.004$&$0.724\pm0.004$ \\ \hline
$\beta$&$0.123\pm0.002$&$0.223\pm0.003$&$0.261\pm0.003$ \\ \hline
$\gamma$&$0.233\pm0.005$&$0.674\pm0.008$&$0.756\pm0.009$ \\ \hline
$\delta$&$-0.217\pm0.015$&$-0.708\pm0.022$&$-0.810\pm0.022$ \\ \hline
$\Tilde{\chi}^2$&$0.43$&$0.75$&$0.97$ \\ \hline
$\Tilde{e}$&$2\%$&$4\%$&$5\%$ \\ \hline
$e_{max}$&$23\%$&$24\%$&$30\%$ \\ \hline
\multicolumn{4}{c}{}\\
\end{tabular}
\caption{Best-fit parameters and related standard errors, for the functional form Eq.~\ref{eq:functionalform} fitted on $Y_{\rm bar}$, $Y_{\rm gas}$ and $Y_{\rm hot}$. The values of the reduced $\chi^2$ ($\Tilde{\chi}^2$), the median and the maximum deviation of the model from the data ($\Tilde{e}$ and $e_{max}$, respectively) are also quoted.}
\label{tab:fitdepletion}
\end{table}

\begin{figure}[hb]
\includegraphics[width=0.49\textwidth]{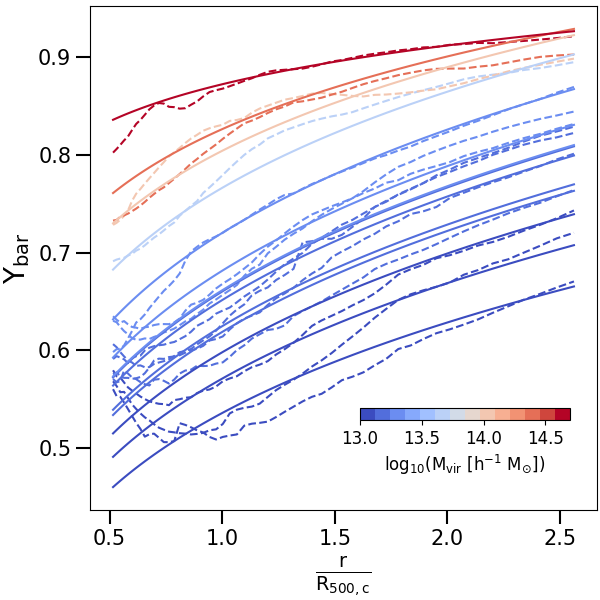}
\caption{Radial profiles of baryon depletion, from $0.5R_{500,\mathrm c}$ up to $2.5R_{500,\mathrm c}$. The dashed lines represent the median profiles in each mass
bin, according to the legend on bottom-right corner. The solid lines are the fit performed according to functional form Eq.~\ref{eq:functionalform}, with the same color scale of median profiles.}
\label{fig:ybar_fit}
\end{figure}

\end{document}